\documentclass[aps,prl,twocolumn]{revtex4}

\usepackage{graphicx}% Include figure files
\usepackage{dcolumn}% Align table columns on decimal point
\usepackage{bm}% bold math

\begin{document}

\title{Atomistic Origin of Urbach Tails in Amorphous Silicon}
\author{Y. Pan}
\author{F. Inam}
\author{M. Zhang}
\author{D. A. Drabold}
\affiliation{Department of Physics and Astronomy, Ohio University, Athens OH 45701, USA}

\date{\today}

\begin{abstract}
Exponential band edges have been observed in a variety of materials, both crystalline and amorphous. In this paper, we infer the structural origins of these tails in amorphous and defective crystalline Si by direct calculation with current {\it ab initio} methods. We find that exponential tails appear in relaxed models of diamond with suitable point defects. In amorphous silicon (a-Si), we find that structural filaments of short bonds and long bonds exist in the network, and that the tail states near the extreme edges of both band tails are are also filamentary, with much localization on the structural filaments. We connect the existence of both filament systems to structural relaxation in the presence of defects and or topological disorder.

\end{abstract}

\maketitle

Urbach first identified exponential (not Gaussian) tails at the edges of optical interband and excitonic transitions in impure crystals more than 50 years ago\cite{urbach53}.   Here we examine the internal structure of models of amorphous Si, and detail the structural origin of the Urbach tail in this archetypal amorphous material.   Because of the wide occurrence of Urbach tails, we suggest that similar identification of such internal structures and analysis of their properties provides a novel path for finding and understanding hidden internal structure in other optimized networks, both electronic and molecular.

At first sight, deriving the Urbach absorption tail, $\rho(E) \propto e^{-|E-E_b|/E_u}$, [here, $\rho(E)$ is the electronic density of states, $E_u$ parameterizes the tail decay into the gap,  and $E_b$ is a band-edge energy], looks easy enough.  Because of the similarity of these exponentials to Arrhenius activated relaxation, intuitive arguments based upon fluctuating potentials in landscape models should yield the desired result, and indeed they do yield exponentials, but the function in the exponential is not linear in $x = E-E_b$; instead, a $\frac{3}{2}$ power is found\cite{halperin,morrel}. A way around this problem is to start with Gaussian distributions of internal strain energies, which leads directly to the exponential with a linear argument (because strain energies are harmonic)\cite{keating}.  The question then arises, what are these strain energies, and how do we identify them, for instance, in the much-studied models of amorphous Si? Thus, the simplest explanation for Urbach tails is to assume an isomorphic correspondence between local distortions (most naturally in bond angles), and a shift from a band edge (larger distortion, larger shift toward midgap). One can easily obtain an exponential tail with assumptions of this kind\cite{dong96}, but this turns out to be flawed: {\it local} (individual) variations in bond length or bond angle may {\it not} be associated with simple energy shifts from the band edge. Thus the structural features of the model giving rise to the Urbach edge must be more complex and nonlocal than this appealing but oversimplified model suggests.

Previous work on large models of a-Si and varied forms of disorder illustrate the qualitative\cite{dong98} and universal nature\cite{ludlam} of the Anderson transition, and reveal that the spatial character of electron eigenstates in the tails are ``islands and filaments": local clusters of charge, often interconnected by filaments\cite{dong98}. The filaments are most evident at the extreme band tail energies; proliferating charge islands become important for energies closer to the mobility edges, as we describe elsewhere \cite{ludlam}. We name the 1-D structures in the tail eigenstates {\it electron filaments} (EF).

In this paper, we use the {\it ab initio} local orbital code SIESTA\cite{siesta}, both for relaxation studies and also for spectral properties (the density of electron states). In all calculations, supercell models are used. In our calculations, we employ a single-zeta-polarized\cite{basis} basis and sample the Brillouin zone at $\Gamma$ for total energies and forces. We extensively sample the Brillouin zone for computations of the density of states.

To unmask the atomistic origin of the Urbach edges, we begin with a 512-atom model of Si in the diamond structure, create two vacancies and relax the network\cite{siesta}. Before relaxation, the model exhibits a sharp band edge like the ideal crystal, with gap states arising from the vacancies. Relaxing the vacancies causes Jahn-Teller distortion \cite{lannoo91} and relaxation involving many atoms, as reported by others \cite{Puska98}. The relaxation is primarily directed along 1-D filaments, propagating away from the defect site. As we illustrate in Fig. 1a, this relaxation leads to an exponential valence tail in the DOS. This observation is consistent with recent ion bombardment experiments, which report the appearance of an Urbach tail well before amorphization \cite{zammit91}\cite{sundari04}, with a characteristic Urbach energy varying in the range  280-370 meV for Si (to be compared with 350 meV for our relaxed vacancy model). It is suggested that the Urbach energies are more sensitive to point like defects created at low ion dose \cite{zhang04}\cite{sorieul04}.

\begin{figure}
%\includegraphics{btail}\\
\resizebox{90mm}{!}{\includegraphics{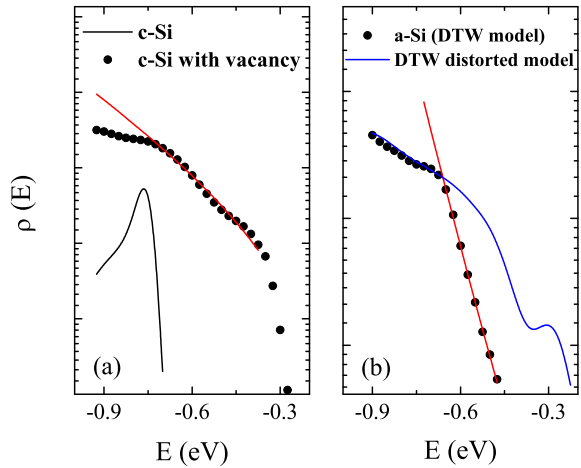}}
  \caption{a) Valence band tails for crystalline Si and Crystal Si with two vacancies. Exponential fit to vacancy model gives Urbach energy $E_U=350 meV$. b) Valence band tails for DTW model and the DTW model with randomly distorted bond lengths. Exponential fit to the relaxed DTW model yields $E_u=107 meV$.}
\end{figure}

In a-Si,  we have examined electronic tails states in nine atomistic (500-4096 atoms) models of a-Si\cite{cmp07}.  The better annealed models \cite{DTW95} \cite{art00} exhibit narrow Gaussian strain distributions of both stretching and bending energies.   We studied a 512-atom model (hereafter named DTW) due to Djordjevic, Thorpe and Wooten \cite{DTW95}, in greatest detail. After relaxation with \cite{siesta}, the density of states is reproduced in Fig. 1b. Exponential decay is evident with an Urbach energy of 107 meV.  As expected, EF are observed\cite{cmp07}. It is of interest that a clear signature of exponential tailing is seen even with a 512-atom model (we have used maximum-entropy techniques to extend such calculations to $10^5$ atoms\cite{dad08}, and such calculations always reveal exponential tails in good quality models). Ion-bombardment experiments (Si on Si) give an Urbach energy of 170-240 meV in fully amorphized samples\cite{zammit91}.

\begin{figure}
\begin{center}
  \includegraphics[width=3.7 in]{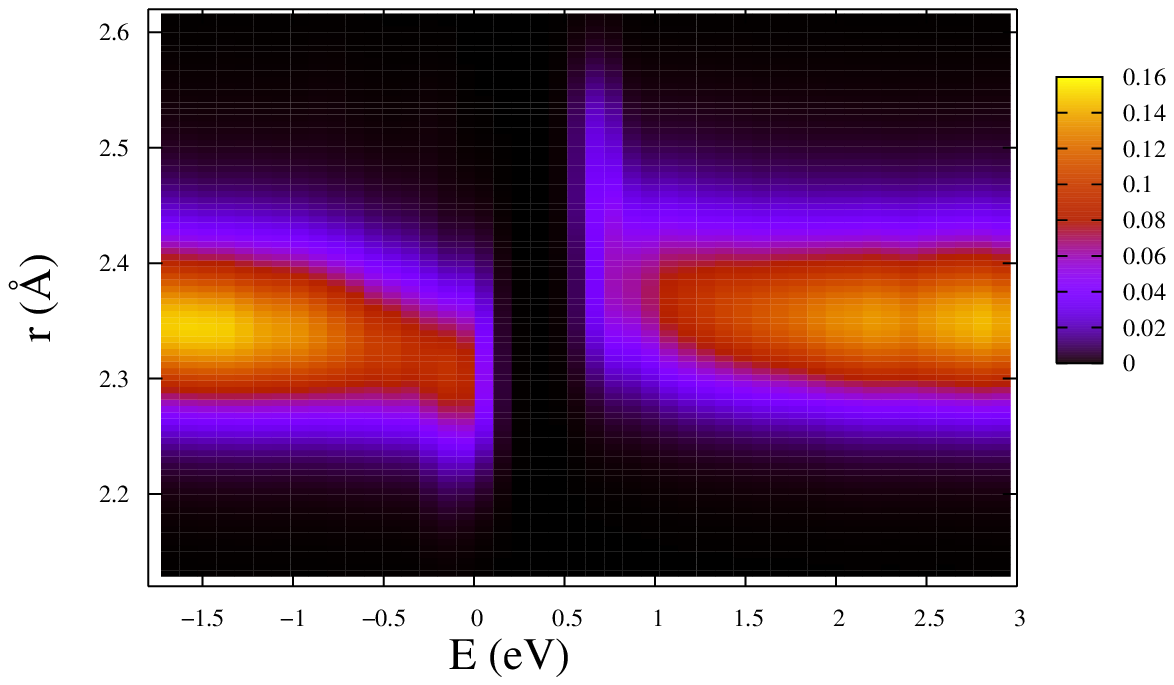}\\
%\resizebox{99mm}{!}{\includegraphics{blqd}}
  \caption{Plot of charge-weighted $Q(n,r)$ (see text), indicating which bondlengths contribute to energy states ($n$) around the gap. Valence (conduction) tail states are derived from short (long) bonds, extended states mostly arise from bond lengths near the mean. The Fermi level lies in the middle of the gap, near $E=0.25eV$.}
  \label{fig2}

  \includegraphics[width=3.7 in]{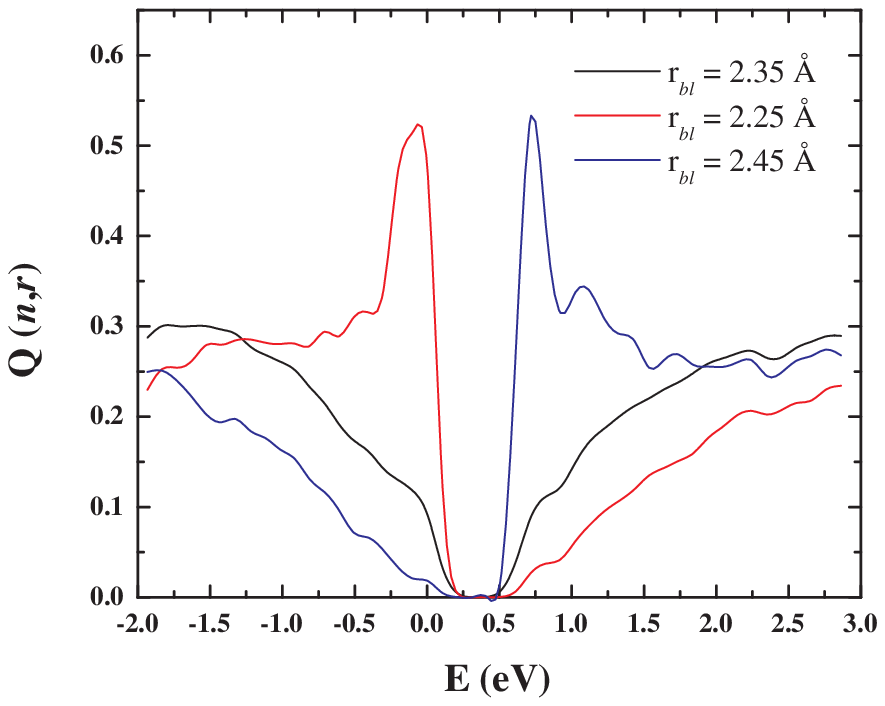}\\
  \caption{Normalized projection of $q(n,r)$ (see text) extracted from Fig. \ref{fig2} for short, mean and long bond lengths. The valence tail states are derived primarily from short bonds, the conduction tail states from long bonds. The mean bond length is near 2.35\AA.}
  \label{fig3}
\end{center}  
\end{figure}

In the DTW and other high-quality models, we have identified {\it topological filaments} (TF) (related to, but distinct from the electron filaments) -- structural patterns that resemble hydrodynamic flow fields analogous to those expected from HelmholtzÕ theorem, namely solenoidal loops (associated with bond bending) and  irrotational strings (or filaments, associated with bond-stretching)\cite{cmp07}.  The DTW model is fully non-crystalline, and relaxed with {\it ab initio} methods.   These TF are shown in Fig. 2 of Ref. \cite{cmp07}, where we also quantify these correlations by computing bond-bond correlation functions for long-long, short-short, and short-long bonds. Because short and long bonds manifestly deviate from the mean structure, we found it interesting to explore their electronic signatures. One means to measure the ``overlap" between EF and TF is to project the electronic eigenvectors in the band tails on the bond lengths by computing the charge weighted mean bondlength $W(n)$ for each eigenstate $n$, defined as $\sum_{i,j}Q(n,r_{ij})/N$, where $Q(n,r_{ij})=q(n,i)q(n,j)r_{ij}$ and $N=\sum_{i,j} q(n,i)q(n,j)$ is the normalization factor, $r_{ij}$ is the bond length of neighboring sites $i$ and $j$, and $q(n,i)$ is the relative weight of energy eigenstate $n$ on atomic site $i$. Calculations were performed\cite{fedders98}, in which it was observed that the valence (conduction) tail is associated with short (long) bonds, albeit without recognition of the existence of TF. In Fig. \ref{fig2} we show the 2-D distribution of the quantity $Q(n,r)$ for states $E_{n}$ around the gap and for all the bond lengths $r$ present in the network. The contribution from normal bond lengths ($2.35$ \AA) is maximum inside the bands and drops to zero at the band edges (Fig. \ref{fig3}). Note the almost perfect symmetry between these projections for the conduction and valence edges. It shows that the band tail states do indeed arise from the structural disorder which, in case of a-Si, appears in the the form of TF. Close examination of TF shows that these structures are nucleated by a primary disorder center where the deviation from the perfect tetrahedral geometry is maximum; this disorder (here, anamolous bond length) decays along the length of the filament. Thus, one can define an Urbach network, whose connectivity arises from the overlap of these TF. Similar to TF we can associate a localization center to EF for the tail states. The overlap between the TF and EF is found to increase for the extreme tail states closest to midgap. As the electron energy varies from Urbach band edge to mobility edge the electron state become less filamentary and more ``island-like"\cite{dong98}.

To further elucidate the role of TF in the formation of exponential band tails, we distort the TF correlations by randomly changing the bond lengths in the filaments without creating coordination defects and without any dramatic change in the overall structure of the network, then we calculate the density of states of this 'distorted' model. We applied this procedure on a 64, 216 atom a-Si model and DTW model. In all cases the band tails extend further in the gap and the localization of conduction tail states decreases considerably with relatively less decrease in valence tail localization. Fig. 1b shows the valence band tail of the DTW model and the 'distorted' model. While the relaxed DTW model shows a clear exponential tail with Urbach energy {\it ca} 107 meV, the distortion in the bond lengths has modified the functional form of the tail from exponential to something more Gaussian. The variation in the properties of band tail states due to the topological fluctuations as we observed in our models is also supported by the effects of external strain on the carrier mobilities in a-Si:H TFT as observed in number of experiments\cite{jones85}\cite{gleskova04}. Applying an external strain would result in the distortion of bond lengths and bond angles, thus would distort the topological correlations effecting the band tails.

\begin{figure}
%  \includegraphics[width=4.2 in]{chains}\\
\resizebox{90mm}{!}{\includegraphics{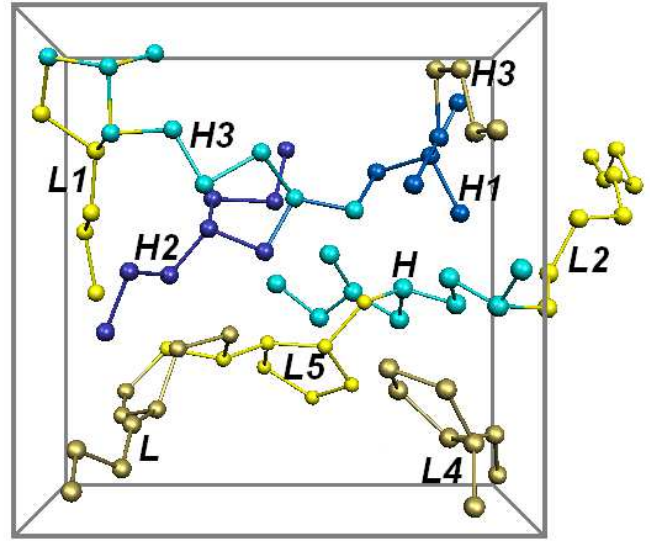}}
  \caption{Electron filaments associated with several states in valence and conduction tails. ``H" refers to highest occupied molecular orbital state, H-1, next state of lower energy and so on, ``L" denotes the lowest unoccupied molecular orbital state, L+1 next highest energy state etc. Valence tail state are selectively localized on short bonds, the conduction tail on long bonds (see text). The valence tail states are filamentary; the conduction tails consist of filaments with occasional rings.}
  \label{fig4}
\end{figure}

Thus a relatively simple picture of the Urbach tail states and their structural origin in a-Si emerges from our work. Near the band edge extrema, the states consist of filaments of charge in the network, that are strongly associated with the structural filaments. In Fig. \ref{fig4}, we show five valence tail and four conduction tail states. Their filamentary structure is clear, albeit including the presence of some even-membered rings. Note further that the valence tail consists primarily of simple filaments, whereas the conduction tail includes a number of rings (a natural consequence of the extended bond lengths of the conduction filaments). Thus one can view the extreme tail eigenstates in a-Si to consist of interpenetrating filaments of long (conduction)  and short (valence) bonds. These filaments are a consequence of structural relaxation. If the long-long or short-short correlations are artificially destroyed the tail is severely affected and the gap is significantly reduced. It is convenient to view the filaments of Fig. \ref{fig4} as something like quasiparticles; the transport and other important electronic properties depend in critical detail on the spatial character of these states and also their coupling to the lattice (phonons). 
 
It is of central interest that the TF and EF overlap, and this overlap has a strong energy dependence (nearer midgap, the higher the overlap). Optical and conductivity measurements are determined primarily by the character of the electron states near the Fermi level, and this is true also for lightly doped materials. Thus, our demonstrated correlation between EF and TF may be interpreted as a link between transport properties and the TF present in a-Si. We have already shown elsewhere that localized and partly-localized deep tail states always have a strong electron-phonon coupling\cite{attafynn}; this paper goes a step further and shows that rather simple 1-D structures are at the root of this large coupling. If one considers the Kubo formula\cite{kubo}, the conductivity is expressible as a sum over transitions between occupied and unoccupied states: transitions between short-bond and long-bond filaments, both with energies specially susceptible to phonons. Hence, one envisions carrier transport as a phonon-driven process among EF. As the TF (correlated with the EF) exhibit distinct structural signatures, the very phonon modes that enhance the transport are determined in part by the TF. 

Recent experiments in the area of cuprate superconductivity employing atomic-resolution tunnelling-asymmetry imaging reveal complex patterns associated with strong uniaxial ``stripe" disorder of the tetragonal basal plane\cite{davis}, as well as patchy disorder associated with superconductive and pseudogaps\cite{hudson}. The latter could create filamentary Urbach tail states which could couple to interlayer dopant charges to form a dopant-based network, as suggested by Phillips\cite{jim}. While there are obviously stark differences between the cuprates and a-Si, the similarity of the toplology of the TF/EF and the STM images is of interest. Finally, we have noted elsewhere that qualitative features of the states near the gap reveal universality.  As eigenenergy is varied from midgap into a mobility edge,  the qualitative evolution of the structure of the states is identical for Anderson models, realistic structural models as discussed here, and even lattice vibrations with mass and spring constant disorder\cite{ludlam}. While the origin of the pseudogap in the cuprates is quite different from the optical gap in a-Si, the similarity in the topology of the states seen in Ref. \onlinecite{davis} and our work hints at the possible existence of TF in the cuprates as well.

In conclusion:  (1) exponential band tails appear even in large relaxed models of Si in a relaxed diamond structure and are associated with the network relaxation induced by the presence of point defects;   (2) we have demonstrated the existence of simple self-correlations between long bonds and short bonds "structural filaments", (3) we show that the tail electron states arise from the structural filaments and compute the energy dependence of this overlap. (4) To the extent that other work has shown that exponential band tails are obtained from a larger DTW model made in a fashion identical to the 512 atom model we employ here, it is a reasonable inference that the Urbach edge in a-Si is due to the existence of the structural filaments. We note that we have so far proven only sufficiency: TF can be the structure underlying the Urbach edge: we have not proven that TF are unique in their characteristic of producing Urbach edges. However, because the TF appear naturally in the best extant models of a-Si, and survive thermal MD simulation and relaxation\cite{cmp07}, we think it reasonable to accept them as the underlying structure responsible for the Urbach edges in a-Si, and possibly other materials.

We wish to acknowledge Dr. J. C. Phillips. His insight and enthusiasm have significantly impacted this work. We appreciate helpful suggestions from Profs. J. R. Abelson and P. M. Voyles. We thank the National Science Foundation for support under grants DMR 0605890 and 0600073 and the Army Research Office under MURI W91NF-06-2-0026. One of us (FI) acknowledges a travel award from the NSF International Materials Institute on New Functionality in Glass (NSF-IMI, DMR-0409588).

\begin{references}
\bibitem{basis}Thus, nine orbitals (one s, three p, five d) are included for each atom.
\bibitem{dad08} We have recently extended the maximum entropy tight-binding order-N method of Ref. \cite{dad93} to realistic models with 10$^4$ and 10$^5$ atoms, kindly provided by Mousseau and Barkema (for their improved WWW method, see Ref. \onlinecite{art00}). In reasonable agreement with the earlier work,  exponential tails were observed at  conduction and valence edges, with $E_u\approx107 meV$.
\end{references}
\end{document}